\documentstyle[twoside,fleqn,espcrc2]{article}

\newcommand{\AmS}{{\protect\the\textfont2
  A\kern-.1667em\lower.5ex\hbox{M}\kern-.125emS}}

\title{Structure Functions, Form Factors, and Lattice QCD
\thanks{Talk presented by B. Andersen-Pugh.}}

\author{Walter Wilcox\address{Department of Physics,\\
        Baylor University,
        Waco, TX 76798, USA}%
        and
        B. Andersen-Pugh\address{Department of Physics,\\
        Luther College,
        Decorah, IA 52101, USA}}

\begin{document}

\begin{abstract}
We present results towards the calculation of the pion electric
form factor and structure function on a
$16^3\times 24$ lattice using charge overlap.
By sacrificing Fourier transform information in two directions, it is seen
that the longitudinal four point function
can be extracted with reasonable error bars at low momentum.
\end{abstract}

\maketitle

\section{INTRODUCTION}

The direct calculation of hadron structure functions by current
overlap techniques is based on the simulation
of the Euclidean hadronic matrix elements
$\langle h({\bf 0}) | T[J_{\mu} ({\bf r},t) J_{\nu} (0)] | h({\bf 0}) \rangle
$,
where $J_{\mu}= q_{u}J^{u}_{\mu}+q_{d}J^{d}_{\mu}$ is the
full electromagnetic current and
$J^{d,u}_{\mu}$ are the $d$, $u$ quark flavor current densities.
Such calculations are likely to be quite
costly, so it seems worthwhile to check the validity of this approach
by less ambitious calculations utilizing the basic current overlap
technique. The longitudinal piece (corresponding to $\mu=\nu=0$ above, where
$J_{0}=i\rho$) of structure functions for
mesonic systems provides such a test\cite{one}. We expect for the pion that
vector dominance should hold, resulting in a known elastic limit
for both the $\rho^{u}\rho^{u}$ (same flavor) and
$\rho^{d}\rho^{u}$ (different flavor) sectors.
This calculation serves to test whether the lattice
size is large enough in space and
time, the size of statistical errors, and other important issues.
Since we use the conserved lattice current, there are also
many useful numerical identities which serve as checks on the calculation.
In constructing these four point functions, we make multiple use of the
sequential source technique\cite{sst} for quark propagators.
Our work so far indicates that the full electromagnetic
amplitude should be attainable for low momentum transfers; however,
we see no indications of inelastic contributions in our results.
In this brief report we will concentrate on issues relating to
systematics and statistical errors.

\section{THE CALCULATION}

\subsection{The Current Overlap Technique}

The structure function can be obtained from\cite{me}
\begin{equation}
Q_{\mu\nu}({\bf q}^2,t) = \sum_{\bf r} e^{-i{\bf q\cdot r}}
						P_{\mu\nu} ({\bf r},t) \label{1}
\end{equation}
where,
\begin{eqnarray}
\lefteqn{P_{\mu\nu}({\bf r},t) =} \nonumber \\
& \sum_{\bf x} \langle \pi^{+}({\bf 0}) |T[
							J_{\mu}({\bf r+x},t) J_{\nu}({\bf x},0)] | \pi^{+}({\bf 0}) \rangle .
\label{2}
\end{eqnarray}
It is the quantity $P_{\mu\nu}({\bf r},t)$ which is directly calculated
in our simulations. We may study the form factor by using the relation
\cite{me2}
\begin{eqnarray}
\lefteqn{Q_{00}({\bf q}^2,t) \stackrel{t>>1}{\longrightarrow}} \nonumber \\
&  \qquad\qquad\quad\frac{(E_{q} + m_{\pi})^2}{4E_{q}m_{\pi}} F^{2}
({\bf q}^{2}) e^{-(E_{q}-m_{\pi})t}.
\label{3}
\end{eqnarray}

For this calculation we utilized a $16^3\times 24$ lattice and
$\beta=6.0$ in the quenced approximation.
We have omitted disconnected quark graph
amplitudes because of the difficulty of
simulating the corresponding correlation functions.
In constructing the $P_{00}({\bf r},t)$ there are three
distinct classes of connected diagrams which contribute;
see Figure 1 where the the effect of the currents is
represented by an \lq\lq {\bf X}".
The different flavor piece (Figure 1(a)) is
calculated by combining quark propagator lines from
source and sink; this involves two quark inversions per
configuration. In addition, the same flavor amplitude
contains both \lq\lq direct" (Figure 1(b))
and \lq\lq Z-graph" (Figure 1(c)) contributions
because of the indistinguishability of the two currents. These
are calculated separately and the results added together;
\begin{figure}
\vskip 65mm
\special{illustration 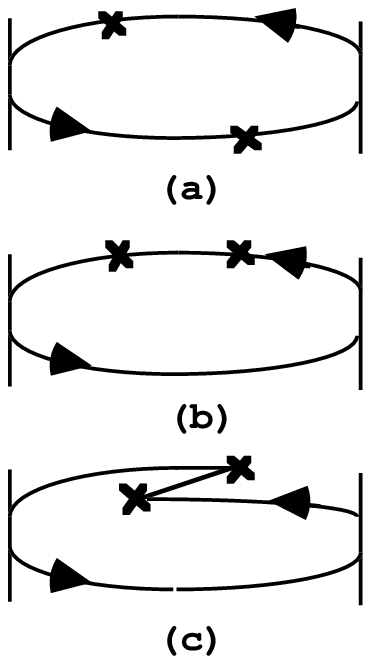}
\caption{Three types of connected current overlap
diagrams: (a) different flavor; (b) same flavor direct;
(c) same flavor Z-graph.}
\label{figure1}
\end{figure}
three additional propagators are needed to do this.

\subsection{Fourier Reinforcement}

It is the ${\bf x}$ sum in equation (\ref{2}) which is difficult to perform
when the currents involve the same flavor because of quark lines
going from $({\bf r}+{\bf x},t)$ to $({\bf x},0)$
(${\bf r}$ and ${\bf x}$ both summed).
However, the statistical error bars on $Q_{00}^{uu}({\bf q}^{2},t)$
were reduced by means of the following strategy.
We use a charge density sheet operator,
\begin{equation}
{\tilde\rho}(z,t) =\sum_{x,y} \rho(x,y,z,t),
\label{4}
\end{equation}
Using this when ${\bf q}=q \hat{{\bf z}}$,
$Q_{00}^{uu}({\bf q}^2,t)$ can be written
(using translational independence) as
\begin{eqnarray}
\lefteqn{Q_{00}^{uu}({\bf q}^2,t)= N_{z}\sum_{z} e^{-iqz}} \nonumber \\
&	\qquad\quad\cdot\langle \pi^{+}({\bf 0}) |T[
							 {\tilde\rho}(z,t) {\tilde\rho}(0,0)] | \pi^{+}({\bf 0}) \rangle,
\label{5}
\end{eqnarray}
where $N_{z}$ is the number of lattice points in the $z$ direction.

\begin{figure}[htb]
\vskip 65mm
\special{illustration 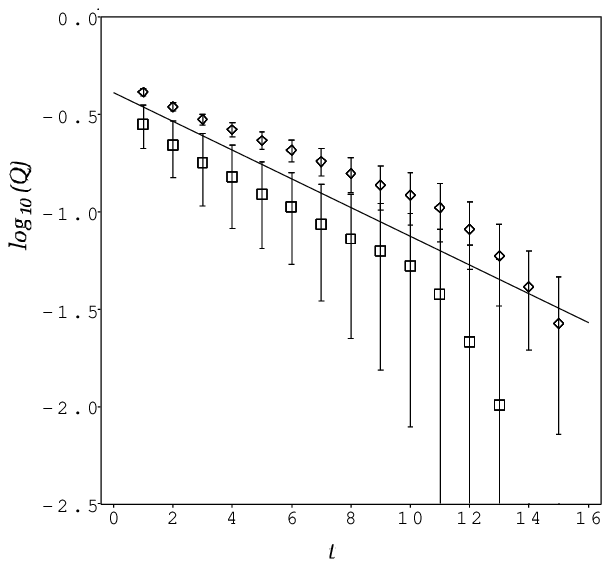}
\caption{Effect of using the extended operator ${\tilde\rho}(z,t)$ at $\kappa
=.154$ on $Q_{00}^{uu}({\bf q}^2,t)$. Diamonds: Fourier Reinforced data;
squares: nonreinforced data. The line
here and in Figure 3 is the expected vector dominance elastic limit.}
\label{figure2}
\end{figure}

Figure 2 illustrates the effect of using this technique
at $q = \frac{\pi}{8}$ and $\kappa = .154$ on $10$ configurations.
The Fourier Reinforced ($\lq\lq FR"$) result
used the ${\tilde\rho}(z,t)$ operator,
whereas the nonreinforced result fixes ${\bf x}$ in (\ref{2})
to a single location\cite{foot}. The two results in Figure 2 are consistent
with
one another, although the $FR$ result is systematically higher.
Both are consistent with a
single exponential behavior (no inelastic part), even at small
time separations between the currents, $t$, although there is
a mysterious bump in the $FR$ result. This could be
a statistical fluctuation or an effect of getting too near
the sink interpolation fields.
It is seen that the uncertainties in $Q_{00}^{uu}({\bf q}^{2},t)$
are significantly reduced in the $FR$ amplitude.

\section{PRELIMINARY RESULTS}

\begin{figure}[htb]
\vskip 65mm
\special{illustration 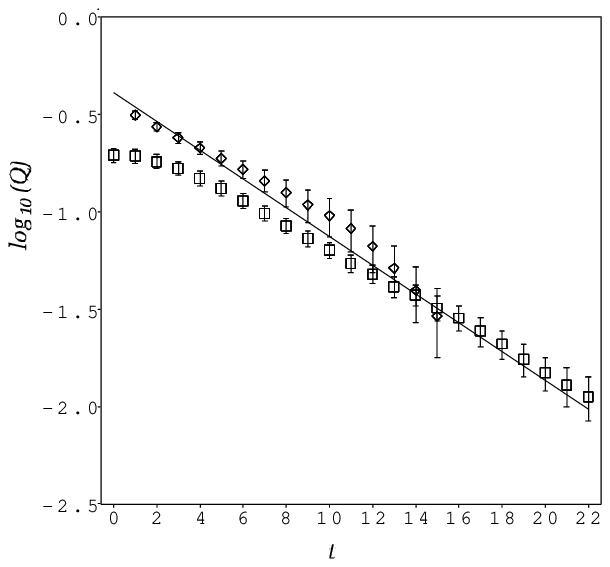}
\caption{Diamonds: full electromagnetic amplitude, $Q_{00}({\bf q}^{2},t)$,
equation (6); squares: different flavor
amplitude, $Q_{00}^{du}({\bf q}^{2},t)$. }
\label{figure3}
\end{figure}
Zero momentum smeared pion interpolation fields are used at source
and sink which are located on the time edges of the lattice
in order to maximize the time extent available for measurements.
Extensive numerical tests were performed on the $Q_{00}^{du}({\bf q}^2,t)$
amplitude to make sure nonvacuum contaminations do not occur; no such effects
were seen. This quantity was reconstructed so that the
two nonlocal, conserved charge densities were centered in time between
the time edges. When measuring $Q_{00}^{uu}({\bf q}^2,t)$,
we used an extended source ($FR$ case) or a point charge source
(non $FR$) on time steps $8$ and $9$. Thus, the maximum time separation
between the two charge operators
is $15$ in the same flavor case but $22$ in the different flavor case.

We find that the the same flavor spatial correlation function
is significantly smaller than the different flavor one. It is
dominated by the direct graph; the
Z-graph contributes only at small time separations and serves to
make the correlation function marginally wider. At small $t$
we see a spatial anisotropy similar to~\cite{negele}.

Figure 3 shows the full pion amplitude
\begin{equation}
Q_{00}({\bf q}^{2},t) = \frac{4}{9}Q_{00}^{du}({\bf q}^{2},t)
+ \frac{5}{9}Q_{00}^{uu}({\bf q}^{2},t),
\label{6}
\end{equation}
again at $q=\frac{\pi}{8}$ for $\kappa=0.154$ compared with
the different flavor piece $Q_{00}^{du}({\bf q}^{2},t)$
(for which the sum in (\ref{2}) is
easy). Using $FR$, the error bars on the full amplitude come
under control. Both results are apparently tending to the vector
dominance elastic limit.

It is possible to measure both elastic and inelastic processes from
our hadronic amplitudes. The various different flavor amplitudes
are especially useful because of their
small error bars. They contain the information to do a
survey of the form factors of all groundstate
nonsinglet mesons.

\section{ACKNOWLEDGMENTS}

This work was partially supported by the National Center for Supercomputing
Applications and the National Science Foundation
under Grant PHY-$9203306$
and utilized the NCSA CRAY systems at the University of
Illinois at Urbana-Champaign.

\end{document}